\documentclass[submission,copyright,creativecommons]{eptcs}
\providecommand{\event}{F-IDE 2021} 
\usepackage{breakurl}             
\usepackage{underscore}           

\usepackage{listings}

\makeatletter
\renewcommand{\verbatim@font}{\footnotesize\ttfamily}
\makeatother

\newcommand{\CodeSymbol}[1]{#1}

\title{How the Analyzer can Help the User Help the Analyzer}
\author{Yannick Moy
\institute{AdaCore}
\email{moy@adacore.com}
}
\def\titlerunning{How the Analyzer can Help}
\def\authorrunning{Y. Moy}
\begin{document}
\maketitle

\begin{abstract}
The automation offered by modern program proof tools goes hand in hand with the
capability to interact with the tool when the verification fails. The SPARK
proof tool tries to help the user by providing the right information, so that
the user can help the tool complete the proof. In this article, we present
these mechanisms and how they work concretely on a simple running example.
\end{abstract}

\section{Introduction}

Program proof is the application of deductive verification techniques to
programs. Industrial acceptability of such tools relies on the high degree of
automation provided by modern automatic provers, in particular SMT solvers,
when the source code is restricted to a suitable language
subset. SPARK~\footnote{\url{https://www.adacore.com/sparkpro}} is an example
of such a FLOSS industrial tool for Ada programs. It is available freely online
as part of GNAT Community
Edition~\footnote{\url{https://www.adacore.com/download}}.

While striving to offer the most automation to our users, we have also
recognized early on the need for interactions when the analyzer cannot complete
the verification automatically. In those cases, the task for the analyzer is to
display to the user the right information that will allow her to provide in
exchange the pieces of information that are required for the analyzer to
complete the verification.

Over the years, we have come up with a variety of solutions to address this
challenge, trying to adapt these solutions to the degree of expertise of the
user. All these solutions came from discussions (usually by email) with
industrial users of SPARK, as part of the support activity that they subscribed
to: a user asks about a puzzling message or a problem with proving a property;
we explain the possible problem and show how this can be investigated; through
discussion we together come up with a way to include part of this explanation
in the messages of the analyzer. We have lots of anecdotal evidence that this
helps users, in the form of positive feedback to our message improvements. We
are presenting these solutions in this article, in the hope that they can be
useful to others, and serve as a basis for better future solutions, as the
challenges presented here are common to most similar analyzers.

\section{The Nurse: Providing First Aid}

At a minimum, the analyzer should help the user understand what is the problem,
using in particular clear messages with precise locations. This is not always
enough, due to the complexity of the detailed program semantics at play which
may elude the user. This difficulty is compounded by the lack of expertise for
novice users of the programming language. The solution we have adopted in SPARK
is to augment the basic message with additional information explaining the
immediate cause of the problem.
Let's consider a procedure \verb|Erase| which replaces every character in a
string with the blank character:

\begin{lstlisting}
   procedure Erase (S : out String) is
   begin
      for J in 1 .. S'Length loop
        S (J) := ' ';
      end loop;
   end Erase;
\end{lstlisting}

\noindent
When running GNATprove (the name of the analysis tool for SPARK) on that code,
it issues the following message:

\begin{verbatim}
strings.adb:6:13: medium: array index check might fail
    6 |         S (J) := ' ';
      |            ^ here
  e.g. when J = 1
        and S'First = 2
  reason for check: value must be a valid index into the array
\end{verbatim}

The main message line contains the precise location of the message, its
severity, and the message text indicating which property might be
violated. Here, this is an array index check, which corresponds to the run-time
check when accessing an array in SPARK. The next two lines are only present in
command-line usage, and help visually locate the issue outside of an IDE. So
far, this is what every tool diagnostic should contain.

The purpose of the remaining lines is to help the user understand the reason
for the message to be issued. The counterexample introduced by ``e.g.'' gives
concrete values of variables which lead to a property violation. Here, when
\verb|J| is one and the string \verb|S| starts at index two (denoted in SPARK
by \verb|S'First|), the assignment attempts to write outside of the bounds of
the array. This is reinforced by the last line, which gives the reason for a
check at this location: assigning into an array in SPARK requires the index to
be within bounds.

These messages are particularly useful for beginners, who may not know that
arrays (and among them, strings) in SPARK may start at other indexes than one,
or who may not realize that assigning into an array involves implicitly a
run-time check that this assignment is within bounds.

The ``reason for the check'' provides additional context which can be useful
also to more advanced users, in cases where the details of the language
semantics requiring a check are more complex. It also facilitates understanding
exactly in which part of a longer expression a check originates, as there might
be many similar checks on the same line.

Only part of the counterexample is displayed above as part of the message. A
counterexample is really a trace inside the subprogram consisting in multiple
program points with values of variables at each program point. In the GNAT
Studio IDE, the user can choose to display the trace with a simple click on a
magnify icon next to the message.

Still in this category of first-aid help, GNATprove strives to present the user
with the smallest sub-property that cannot be proved, in cases where the
property to prove is a conjunction of smaller sub-properties or a universally
quantified property. This is particularly useful when attempting to prove the
precondition of a call or a loop invariant, which typically are a conjunction
of sub-properties. This requires GNATprove to split unproved properties further
until a leaf sub-property is not proved.

\section{The Investigator: Looking for Probable Cause}
\label{sec:the-investigator}

Once the user understands the immediate cause of the problem, the next step is
to understand its root cause. Similar to the use of backtrace when debugging a
program, the analyzer should provide information on the context of the problem
that helps identify the missing link in the chain of deductions that the user
does in her head. In particular, programmers rely on operational semantics to
understand program executions, while the analyzer relies on axiomatic semantics
which may abstract away crucial details of the operational semantics. The
solution we have adopted in SPARK is to output additional information related
to the problem when there is a chance that it might have been overlooked by the
user.
Let's fix our implementation of \verb|Erase| by iterating over the range of
string S:

\begin{lstlisting}
   procedure Erase (S : out String) is
   begin
      for J in S'Range loop
        S (J) := ' ';
      end loop;
   end Erase;
\end{lstlisting}

\noindent
and let's add a contract to \verb|Erase| specifying in a postcondition that all
characters in \verb|S| should be blank on return:

\begin{lstlisting}
   procedure Erase (S : out String)
     with Post => All_Blanks (S);
\end{lstlisting}

\noindent
which is defined recursively over the range of \verb|S|:

\begin{lstlisting}
   function All_Blanks (S : String) return Boolean is
     (if S = "" then True
      else S (S'First) = ' '
        and then All_Blanks (S (S'First + 1 .. S'Last)));
\end{lstlisting}

\noindent
When running GNATprove on that code, it issues the following messages:

\begin{verbatim}
strings.ads:9:19: medium: postcondition might fail
    9 |     with Post => All_Blanks (S);
      |                  ^~~~~~~~~~~~~
  possible fix: loop at strings.adb:5 should mention S in a loop invariant
    5 |      for J in S'Range loop
      |                       ^ here
\end{verbatim}

The three last lines are new compared to the messages in the previous
section. GNATprove points here to a possible cause for the failure to prove the
postcondition, which is that the loop in \verb|Erase| has no loop invariant. It
comes to this conclusion by looking at the variables which are mentioned
(explicitly or implicitly) in the property to prove, here only \verb|S|, and
traverses the code in reverse execution order from the program point where the
property should hold. During this traversal, it correctly identifies here that
the loop modifies \verb|S| without specifying how those changes impact the
value of \verb|S| inside a loop invariant, which is a likely cause for not
being able to prove the postcondition.

A common pitfall of program proof is the frame problem. For automatic provers
to be able to reason about formulas that represent the program semantics, these
formulas necessarily must encode small parts of the whole program
semantics. Thus GNATprove defines a frame for each property to check, that only
presents a subset of the information available in the program, abstracting in
particular subprogram calls as the corresponding subprogram contract and loop
iterations as the loop invariant for that loop. When the user did not write a
contract for a subprogram, or did not write a loop invariant for a loop,
GNATprove may still be able to analyze the corresponding call/loop precisely by
inlining the call or unrolling the loop. But this is not always the case, which
raises the question of how these internal tool decisions are communicated to
the user. Because this kind of information was a source of confusion for
beginners, GNATprove only outputs it when instructed to do so with the switch
\verb|--info|, in which case it issues here the following messages:

\begin{verbatim}
strings.adb:5:24: info: cannot unroll loop (too many loop iterations)
strings.ads:6:18: info: expression function body not available for proof
                        ("All_Blanks" might not return)
\end{verbatim}

The first line informs the user that the loop in \verb|Erase| could not be
unrolled (hence it requires a loop invariant) because doing so would require
too many loop iterations. Indeed, a string in SPARK is an array over the range
of positive (32-bits) integers, a range much too large to unroll the loop.

The second line informs the user of another problem here: although function
\verb|All_Blanks| was defined as an expression function (a purely functional
expression for a function, which can readily be translated into an axiom for
proof), its defining expression cannot be used for interpreting the
postcondition here. Indeed, \verb|All_Blanks| is defined recursively, which
makes it possible that it does not return on some inputs. In such cases, it
would be unsound for GNATprove to treat its defining expression as an axiom in
proof, which prevents using it here. There are multiple ways to solve this
problem, either by providing a subprogram variant in order to prove
termination, or by expressing \verb|All_Blanks| differently without recursion.

Still in this category of probable cause, GNATprove can attempt to detect
inconsistencies in specifications or code, by trying to prove that the logical
context for a given branch in the specification or the program entails the
False proposition. As this involves additional calls to automatic provers,
hence has an impact on running time, this is only done when the user chooses to
do so with the switch \verb|--proof-warnings|.

\section{The Magician: Suggesting a Possible Fix}

The ultimate goal of interactivity is to suggest a possible fix to the user, in
those (alas, few!) cases where it is possible, either because some information
is clearly missing, or because a faulty pattern can be recognized. Our
experience with SPARK has shown a few such cases where the analyzer just stops
short of fixing the code itself.
Let's add a loop invariant to the loop in \verb|Erase|:

\begin{lstlisting}
   procedure Erase (S : out String) is
   begin
      for J in S'Range loop
        S (J) := ' ';
        pragma Loop_Invariant (for all K in S'First .. J => S (K) = ' ');
      end loop;
   end Erase;
\end{lstlisting}

\noindent
and reimplement \verb|All_Blanks| without recursion as follows:

\begin{lstlisting}
   function All_Blanks (S : String) return Boolean is
   begin
      for J in S'Range loop
         if S (J) /= ' ' then
            return False;
         end if;
      end loop;
      return True;
   end All_Blanks;
\end{lstlisting}

\noindent
When running GNATprove on that code, it issues the following messages:

\begin{verbatim}
strings.ads:7:19: medium: postcondition might fail, cannot prove All_Blanks (S)
    7 |     with Post => All_Blanks (S);
      |                  ^~~~~~~~~~~~~
  possible fix: you should consider adding a postcondition to function All_Blanks
  or turning it into an expression function
\end{verbatim}

The poscondition of \verb|Erase| still cannot be proved. GNATprove this time
has a more precise suggestion for the solution, which is to add a postcondition
to \verb|All_Blanks| or to turn it into an expression function. Indeed,
GNATprove handles differently regular functions, which may themselves contain
imperative constructs like loops, and so-called expression functions, which can
be readily interpreted in logical terms. Another option would be to add a
postcondition to \verb|All_Blanks|, which the message also mentions.

A similar case where GNATprove can suggest a precise fix to the user relates to
the choice made in GNATprove to prove the absence of run-time errors inside
preconditions independently from calling contexts. So if a subprogram has the
expression (A and B) as a precondition, neither the evaluation of A nor the
evaluation of B should lead to an error. In many cases though, evaluating B
might require that A evaluates to True, and in such cases the precondition
should be expressed using the shorthand connective ``and then'' as (A and then
B). GNATprove detects cases where the user could have used ``and then'' instead
of ``and'' in preconditions to protect against errors, and suggests this
possible fix.

Another such case is the well-known misuse of a conditional inside an
existential quantification, which beginners are almost certain to be bitten by
at some point. When it finds such a syntactic construct (for some X $=>$ (if P
then Q)) which will evaluate to True whenever P is False, GNATprove issues a
warning suggesting the likely fixes:

\begin{verbatim}
file:line:column: warning: suspicious expression
  did you mean (for all X => (if P then Q))
  or (for some X => P and then Q) instead?
\end{verbatim}

While this degree of feedback to the user is highly desirable, it is hard to
produce in general, outside of the specific common cases described above.

\section{The Surgeon: Looking at the Innards}

This exploration would not be complete if we did not present the way for users
to look at the innards of a Verification Condition, in cases where the analyzer
did not present the information needed to understand the problem. Note however
that the preferred means to investigate such unproved properties in SPARK is
through so-called auto-active verification, where the user states intermediate
properties through ghost code (assertions and lemmas).
Let's define \verb|All_Blanks| as an expression function whose body is the same
universally quantified property that we wrote in the loop invariant:

\begin{lstlisting}
   function All_Blanks (S : String) return Boolean is
     (for all J in S'Range => S (J) = ' ');
\end{lstlisting}

When running GNATprove on that code, it proves the postcondition of
\verb|Erase|, but issues messages related to possible reads of uninitialized
data (which were in fact issued on previous versions of the example), of the
form:

\begin{verbatim}
file:line:column: "S" might not be initialized
\end{verbatim}

The reason is that, by default, GNATprove checks correct data initialization by
data flow analysis instead of proof, which is not sufficient here to prove that
\verb|S| is progressively initialized in the loop, which ends with \verb|S|
being completely initialized, and that only the initialized part of \verb|S| is
read in the loop invariant. The solution here is to indicate to GNATprove that
we want it to treat \verb|S| as partially initialized, and to use proof to
demonstrate correct initialization before use:

\begin{lstlisting}
   procedure Erase (S : out String)
     with Post => All_Blanks (S),
          Relaxed_Initialization => S;
\end{lstlisting}

The effect is not immediately visible, as GNATprove keeps issuing messages
about possible reads of uninitialized data. Ignoring for a moment that the
User's Guide explains how to deal with such cases, we can try to understand by
ourselves the underlying model used in proof to deal with
initialization. Through a contextual menu, we can start manual proof on one of
the unproved check, which opens multiple panels in the
IDE~\cite{dailler:hal-01936302}: a panel showing the proof tree (consisting in
the tree of transformations and sub-goals), a panel displaying the current goal
with names translated to reflect source code variable names (with hypotheses
and conclusion), and a panel to enter commands to interact with the tool. After
introducing quantified variables and hypotheses with the command
\verb|split_vc|, the goal looks like this:

\begin{verbatim}
goal def'vc : __attr__init (get2 S _f) = True
\end{verbatim}

\noindent
We can display the definition of \verb|get2|:

\begin{verbatim}
> print get2
function get2 (f:'a -> 'b) (x:'a) : 'b = f \@ x
\end{verbatim}

This is the application of a map representing the string to an index in order
to get the corresponding element. Thus, the Verification Condition here looks
at some attribute \verb|__attr__init| representing the initialization status of
this value, which should be the boolean True to denote that the value has been
initialized. We can search for occurrences of \verb|__attr__init| in the
background theory encoding the program semantics and in the hypotheses encoding
the subprogram execution, using the command \verb|search|:

\begin{verbatim}
> search __attr__init
type character__init_wrapper =
  | character__init_wrapper'mk (rec__value:character) (__attr__init:bool)

function character__init_wrapper___attr__init__projection (a1:
  character__init_wrapper) : bool = __attr__init a1
\end{verbatim}

Here it returns elements of the background theory which allow to attach an
initialization value to a character using type constructor
\verb|character__init_wrapper'mk| and to retrieve the corresponding value from
the pair using function
\verb|character__init_wrapper___attr__init__projection|. And indeed the
constructor is used to define what it means to initialize a character in
function \verb|to_wrapper|:

\begin{verbatim}
function to_wrapper (x:character) : character__init_wrapper =
  character__init_wrapper'mk x True
\end{verbatim}

\noindent
which is used in one of the hypotheses to indicate that \verb|S(J)| is
initialized after the assignment inside the loop:

\begin{verbatim}
H1 : S = set2 S1 J (to_wrapper o)
\end{verbatim}

So we only get that the element at the current index \verb|J| of the string
\verb|S| is initialized, which is not sufficient here. What we need is to
specify in a loop invariant that elements up to the current index have been
initialized, using the attribute \verb|‘Initialized| in SPARK:

\begin{lstlisting}
         pragma Loop_Invariant (for all K in S'First .. J => S (K)'Initialized);
\end{lstlisting}

With that additional loop invariant, everything is proved about \verb|Erase|,
including the additional postcondition that \verb|S| is fully initialized on
return:

\begin{lstlisting}
   procedure Erase (S : out String)
      with Post => All_Blanks (S) and then S'Initialized,
           Relaxed_Initialization => S;
\end{lstlisting}

As visible from this example, looking at the innards of proof requires some
expertise which can only be acquired with time, to understand the mapping from
source language constructs to logical encodings. The use of source variable
names like\verb|S| and \verb|J| in the Verification Condition presented to the
user is a first step towards more systematic roundtrip translation into
constructs at the source code level, to facilitate this understanding.

\section{Related and Future Works}

Counterexamples are the main feature discussed in the context of interacting
with program proof tools~\cite{Cok2010,goues11sefm}. Ideally, a counterexample
captures in an understandable executable trace why a property cannot be proved,
by exhibiting a consistent example where the property does not hold. In
reality, after multiple person-year efforts to develop and improve
counterexamples in the context of SPARK~\cite{dailler:hal-01802488}, there is
still much to be desired here.

On the one hand, counterexamples are very valuable to beginners, to point at
implicit assumptions they might have about the language or program semantics as
well as misunderstandings about the way program proof in general or the
specific SPARK proof tools work. This is particularly valuable for proving
implicit properties of programs like absence of run-time errors, as programmers
are not used to thinking about non-happy paths: How can a signed integer
division overflow? How can a floating-point multiplication between positive
values give zero as a result? On the other hand, counterexamples can be
confusing when they are either spurious because they only reflect a limitation
of the underlying provers (e.g. regarding non-linear arithmetic) or they do not
represent a possible execution but only a limitation of the approach
(e.g. related to the frame problem).

To reduce the possibility of confusion, we decided recently in SPARK to only
enable counterexamples by default at higher levels of proof effort (when
multiple provers are invoked for more than a few seconds per Verification
Condition), so that counterexamples are generated in fewer cases and only in
cases deemed difficult to prove. In parallel, we are working on using symbolic
execution to verify the execution trace represented by a counterexample, in
order to filter out spurious counterexamples and to better label the underlying
error as a property violation or as a frame problem. Previous work on symbolic
or concrete execution have shown benefits for better exploiting
counterexamples~\cite{muller11fm,jacobs11nfm,Hentschel2016,christakis16tacas,petiot16tap}.

Our experience with SPARK is that engineers have difficulties understanding the
axiomatic semantics on which program proof is based. Their main mode of
reasoning about programs is through whole-program operational semantics, which
is supported by tools such as debuggers, fuzzers, profilers, etc. It is thus
critical to provide tool feedback highlighting gaps between the two semantics
which may explain why a property cannot be proved, such as the messages that we
presented in the section~\ref{sec:the-investigator}. This is an area where we
will continue to look for ideas and improvements in the coming years.

\section{Conclusion}

Program proof is intrinsically an interactive effort between a human and a
machine, as complete automation is not achievable. Thus, we are doomed to hit
the so-called Left-Over Principle of automation~\cite{leftover,leftover2},
which is that tasks that are not automated are precisely tasks where humans may
not fare well either, because they are very infrequent or complex. More
generally, the general understanding of the cooperation between the user and
the tool in program proof could be improved, which may require the help of
cognitive science~\cite{tenchallenges}. And notations used to convey
information to the tool could also benefit from the point of view of cognitive
science~\cite{cognitive}. Just looking at the specific issue of tool messages,
the lessons learned from research on compiler error messages have direct
implications for the design and implementation of program proof
tools~\cite{comperr}. In this article, we presented the current state of such
machine-to-human interactions in the SPARK technology, in the hope that it can
trigger interesting human-to-human interactions in the community around proof
tools.

\nocite{*}
\bibliographystyle{eptcs}
\bibliography{article}
\end{document}